\documentstyle[12pt,epsfig,addforfigure]{iopart}
\newcommand{\gsim}{\:\raise -4pt\hbox{$\stackrel{\textstyle >} {\sim}$}\:}
\begin{document}

\title[]{Measurement of open charm production in $d$+Au collisions at $\sqrt{s_{_{NN}}}$=200 GeV}
\author{An Tai\footnote[1]{atai@physics.ucla.edu, for complete author list, see Appendix of these proceedings.} , for STAR Collaboration
}

\address{Dept. of Physics, University of California at Los Angeles, Los Angeles, CA 90095}

\begin{abstract}
We present the first comprehensive measurement of $D^{0}, D^{+}, D^{*+}$ and their charge conjugate states at mid-rapidity 
in $d$+Au collisions at $\sqrt{s_{_{NN}}}$=200 GeV using the STAR TPC.  The directly 
measured open charm  multiplicity distribution 
covers a broad transverse momentum region of 0$<p_{T}<11$ GeV/$c$. The measured $dN/dy$ at mid-rapidity for $D^{0}$ is
$0.0265\pm 0.0036 (stat.) \pm 0.0071 (syst.)$ and  the measured $D^{*+}/D^{0}$ and $D^{+}/D^{0}$ 
ratios are approximately equal with a magnitude of $0.40\pm 0.09(stat.) \pm 0.13(syst.)$. The total $c\bar{c}$ 
cross section per nucleon-nucleon collision extracted from this study is $1.18 \pm 0.21(stat.) \pm 0.39(syst.)\  $mb. 
The direct measurement of open charm production
is consistent with STAR single electron data. This cross section is higher than expectations from PYTHIA and other pQCD calculations.
The measured $p_{T}$ distribution is harder than the pQCD prediction using the Peterson fragmentation function.
\end{abstract}




\section{Introduction}
Study of open charm production provides valuable test of perturbative QCD (pQCD) predictions for heavy quark
production~\cite{heavybig}. There are significant uncertainties in the leading order (LO) and next-to-leading order (NLO) pQCD calculations of
charm quark production depending on choices of the charm quark mass, the renormalization/factorization 
scale and the parton distribution function (PDF)~\cite{vogt1}. Therefore, the open charm measurement
provides a constraint on the parameters used in the pQCD calculations. The energy dependence of the $c\bar{c}$ production cross section
is an important issue. Previously, 
most measurements of open charm production in hadronic interactions were from 
fixed-target experiments at lower energy ($\sqrt{s}\le $63 GeV). Only recently, CDF reported new results on open charm
production in $p\bar{p}$ collisions in the high $p_{T}$ region from the collider experiment 
at $\sqrt{s}$=1.96 TeV~\cite{cdf}. The open charm cross section has been studied in Au+Au collisions
at $\sqrt{s_{_{NN}}}$=130 GeV indirectly through measurements of single electrons from 
the open charm semi-leptonic decay~\cite{phenixele}.

The charm quark is generally believed to be produced primarily through gluon fusion 
($gg\rightarrow c\bar{c}$) in parton-parton hard scattering due to its heavy mass,
which makes the charm quark an important probe to the properties of matter formed in the
early stage of relativistic heavy ion collisions. Theoretical calculations predicted
that the open charm cross section in AA collisions would be significantly enhanced with respect 
to purely hadronic production~\cite{openenhance} as a consequence of quark-gluon
plasma (QGP) formation. Some studies have already attributed the 
increased yield of dileptons in the intermediate mass region reported by 
NA50 to open charm enhancement~\cite{na50}. 
On the other hand, the open charm momentum spectrum may also be modified in AA collisions 
due to final state interactions such as parton energy loss in medium. However, 
it is predicted that energy loss
of a charm quark should be considerably smaller compared to a light quark as a consequence of
the dead-cone effect, leading to an enhanced $D/\pi$ ratio as a function of $p_{T}$~\cite{deadcone}. 

It is found that the relative fraction of the charm quark hadronization into different open charm
species is nearly the same in elementary particle collisions 
($\gamma p$, $e^{+}e^{-}$, $ep$ etc)~\cite{universal}. Could the charm quark hadronization be 
modified in AA collisions? A calculation based on the Statistical Coalescence 
Model~\cite{stat} does show that
ratios of open charm species in AA collisions would be different from those in the elementary
collisions.  Furthermore, at RHIC, the predicted $J/\psi$ suppression has
to be investigated with respect to the total open charm cross section since the 
Drell-Yen cross section, which is the reference of the $J/\psi$ suppression at SPS, is small
in comparison with the open charm cross section at RHIC. In order to investigate the 
outstanding physics
issues mentioned above, it is crucial to study open charm production in a less
complicated system, where the nuclear effects, like QGP formation, parton energy loss, are
not expected. The $d$+Au experiment at RHIC provides  such an essential reference.

\section{Data Analysis}
The main detector involved in this data analysis is the STAR Time Projection Chamber (TPC) 
in a 0.5 T solenoidal magnetic field.
A Zero Degree Calorimeter (ZDC) in the Au beam direction was used for the minimum bias trigger by
requiring at least one beam-rapidity neutron in this ZDC. This trigger accepts 95$\pm3$\% of the $d$+Au
hadronic cross section. For a detail description of the STAR detector, see ~\cite{tpc}. 
In this study, about 15 million $d$+Au minbias events with primary vertex position in
beam direction within $\pm 75\ $ cm   around the TPC center were analyzed. The primary vertex reconstruction
efficiency is 93$\pm$ 1\% of triggered minimum bias events. The correlation between the measured momentum and the ionization energy loss (dE/dx) of charged particles 
in the TPC gas were used for particle identification. The measured $\langle dE/dx \rangle$ can be  well described
by the Bethe-Bloch function smeared with a resolution of width $\sigma$. 

The open charm are reconstructed through their hadronic decay channels 
(and their charge conjugates): 
$D^{0}\rightarrow K^-\pi^+$ (BR=3.8\%), $D^{0}\rightarrow K^-\pi^+\rho^0 (BR=6.2\%)\rightarrow 
K^-\pi^+\pi^+\pi^-$, $D^{+}\rightarrow K^-\pi^+\pi^+$(BR=9.1\%) and
$D^{*+}\rightarrow D^{0}\pi_s^+(BR=68\%)\rightarrow K^-\pi^+\pi^{+}_{s}$. Event mixing technique~\cite{mixing} was used
to reconstruct open charm mass peaks.  The measured charged multiplicity distribution in the region of $|\eta|<0.5$ was divided into 40 bins and the primary vertex position distribution was divided into 15 primary vertex bins. Each event was  mixed with 5 other events within the same
multiplicity bin and primary vertex bin. Tracks within 2 $\sigma$ of the pion Bethe-Bloch curve and 3 $\sigma$ of the kaon Bethe-Bloch curve were selected. In the momentum region of
$p_{T}\gsim $ 0.6 GeV/$c$ where the kaon and pion dE/dx bands merge, the nominal kaon and pion masses were
assumed in turn for each track. It was also required that 
the closet approach distance of a track to the primary vertex, $dca$, be smaller than 1.5 cm with exception of 
the soft pions, 
$\pi_{s}$, from  $D^{*}$ decays where $dca<3$ cm was used.

In the following subsections, 
We will discuss the analysis method for each decay channel.
\subsection{$D^{0}\rightarrow K^-\pi^+ ~(+c.c.)$}
For this analysis only tracks within $|\eta|<1$ were accepted. The transverse momentum and 
total momentum cut of a track were between 0.2 to 10 GeV/$c$ and 0.3 to 10 GeV/$c$, respectively. 
 The resulting
invariant mass distribution after event-mixing background subtraction was shown in Fig.(\ref{d00}a), where a clear $D^{0}$ signal was seen alone with
a residual background distribution. We fit the histogram with a Gaussian function for signal 
plus a linear background.
From the fit, the mass and width ($\sigma$) of the $D^{0}$  signal was found 
to be 1.863$\pm$ 0.003 GeV/$c^2$ and 13.8$\pm$2.8 MeV/$c^2$, respectively. 
\subsection{$D^{0}\rightarrow K^-\pi^+\rho^0 ~(+c.c.)$}
\begin{figure}[htb]
\begin{minipage}[h]{76mm}
\epsfig{file=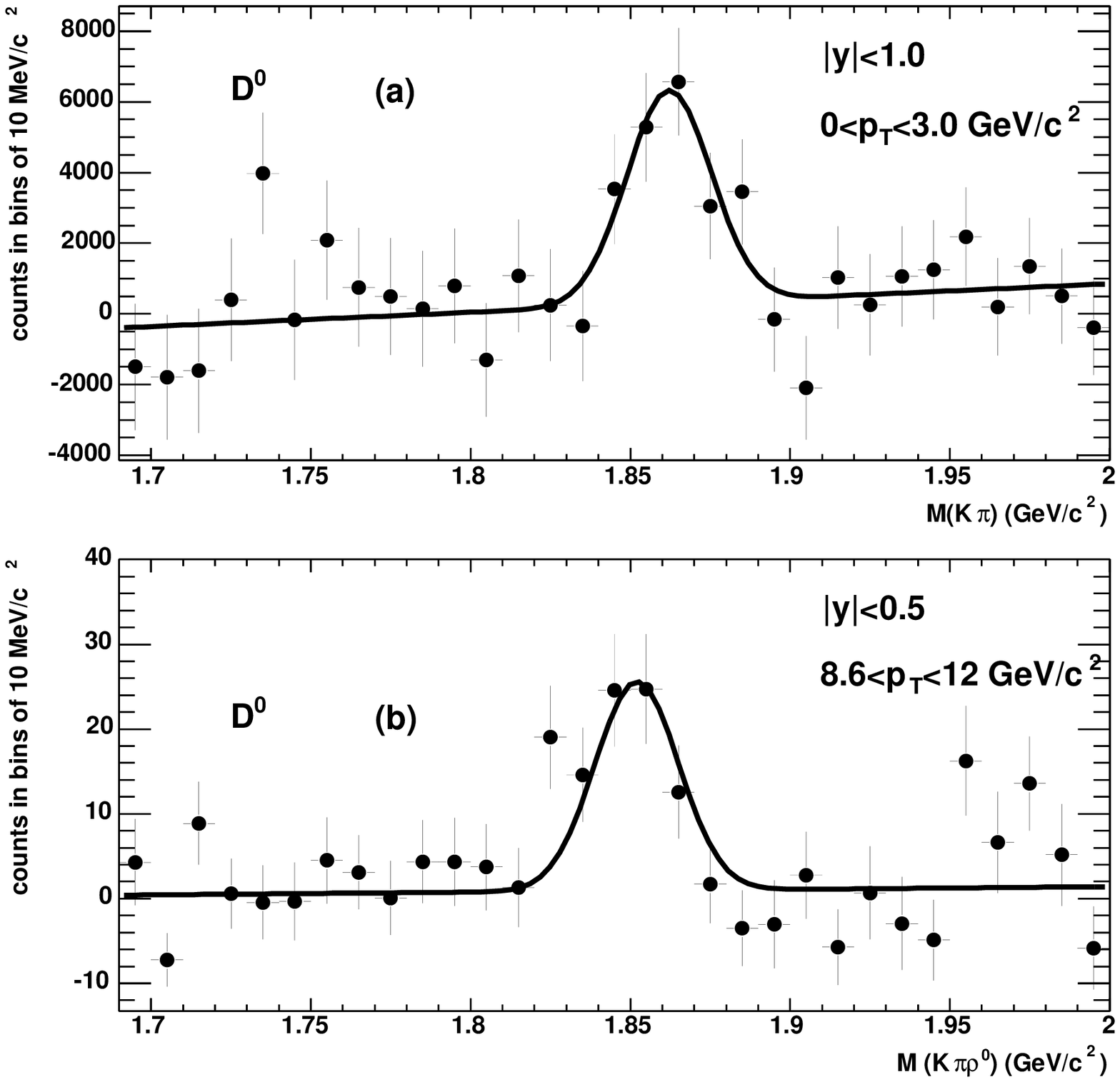,width=7.8cm, height=6. cm} 
\caption{The $D^{0}$ signal reconstructed from $D^{0}\rightarrow K^-\pi^+ ~(+c.c.)$ (a) and 
the $D^{0}$ signal reconstructed from $D^{0}\rightarrow K^-\pi^+\rho^{0} ~(+c.c.)$ (b). }
\label{d00}
\end{minipage}
\hspace{\fill}
\begin{minipage}[h]{76mm}
\epsfig{file=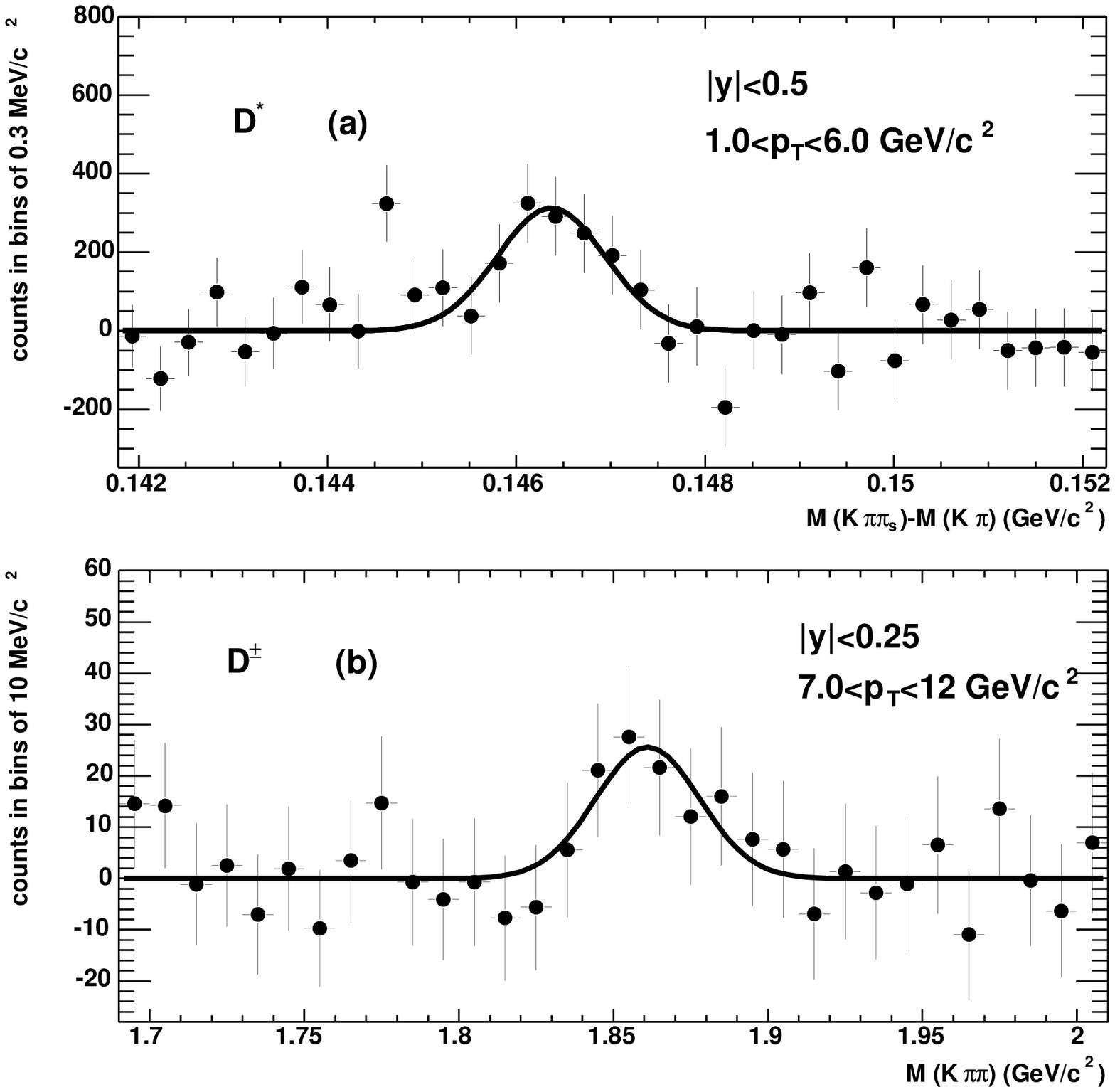,width=7.8cm,height=6. cm}
\caption{The $D^{*}$ signal reconstructed from $D^{*}\rightarrow K^-\pi^+\pi^+_{s} ~(+c.c.)$ (a) and
the $D^{+}$ signal reconstructed from $D^{+}\rightarrow K^-\pi^+\pi^+ ~(+c.c.) (b).$}
\label{d02}
\end{minipage}
\end{figure}

An independent analysis for $D^0\rightarrow K^-\pi^+\rho^0$ was performed where only tracks within 
$|\eta|<1.5$ were accepted. The transverse momentum and 
total momentum cut of a track were set between 0.3 to 10 GeV/$c$. The analysis is similar to that for
$D^{0}\rightarrow K^-\pi^+$ except that  additional  $\pi^+$ and $\pi^-$ candidate tracks, if their 
invariant mass satisfied 0.62 GeV/$c <M(\pi^+\pi^-)<0.86$ GeV/$c$,  were combined 
with the selected $K^-$ and $\pi^+$ candidate tracks to form a $D^{0}$ candidate with invariant mass
M($K^-\pi^+\pi^+\pi^-$). The mass plot after event mixing  background subtraction is shown in Fig.(\ref{d00}b).
We fit the histogram  with a Gaussian function for signal 
plus a linear background. From the fit, the mass and width ($\sigma$) of the $D^{0}$  signal were found 
to be 1.852$\pm$ 0.005 GeV/$c^2$ and 13.6$\pm$4 MeV/$c^2$, respectively. 

\subsection{$D^{*+}\rightarrow D^{0}\pi^{+}_{s} ~(+c.c)$}
The track cuts for kaon and pion candidates from $D^{0}$ decay in the $D^{*+}$ analysis were the same as those
used in the $D^{0}\rightarrow K^-\pi^+\rho^0$ analysis. However, special treatments were needed for the soft
pion daughter, $\pi_{s}^+$, which has an average momentum of about 50 MeV. For the soft pion candidates 
it was required that the transverse momentum and 
total momentum cut be set between 0.1 to 1.0 GeV/$c$ and the ratio of the $D^{0}$ to  $\pi_{s}$ momentum, 
$p(D^{0}$)/$p(\pi_{s})>9.0$. First, the invariant mass of kaon and pion candidate tracks, $M(K\pi)$, 
was calculated and the  $D^{0}$ candidate which satisfied
1.82$<M(K\pi)<$1.90 GeV/$c^2$ were kept. A soft pion candidate, with a charge opposite to
that of the kaon candidate, was then combined with the $D^{0}$ candidate to form a $D^{*}$ candidate with invariant
mass $M(K\pi\pi_{s})$. Fig.(\ref{d02}a) shows the distribution of the mass difference, 
$\Delta M=M(K\pi\pi_s)-M(K\pi)$, for $D^{*\pm}$ candidates after all cuts. A clear signal was seen around the
nominal value of $M(D^{*\pm})-M(D^{0})$. 
We fit the mass distribution  with a sum of a Gaussian function describing the signal and a linear background.
From the fit, 
the mass difference and width ($\sigma$) of the $D^{*+}$ signal were found 
to be 146.37$\pm$0.15  MeV/$c^2$ and 0.57$\pm$0.16 MeV/$c^2$.
\subsection{$D^{+}\rightarrow K^-\pi^+\pi^{+} ~(+c.c)$}
The track cuts used for the $D^{\pm}$ analysis were the same as those for the $D^{*}$ analysis  with the
exception that no 
special treatment for the pion candidate was needed. Fig.(\ref{d02}b)
shows the $M(K\pi\pi)$ distribution for the  $D^{\pm}$ candidates after all the cuts. A clear signal was seen at the
nominal value of $M(D^{\pm})$. The mass distribution was fit to a sum of a Gaussian function describing 
the signal and a linear background. From the fit, 
the mass and width ($\sigma$) of the $D^{*+}$ signal were found 
to be 1.861$\pm$ 0.008 GeV/$c^2$ and 16.9$\pm$4.6 MeV/$c^2$.



\section{Results and Discussions}
\begin{figure}[htb]
\begin{minipage}[h]{76mm}
\epsfig{file=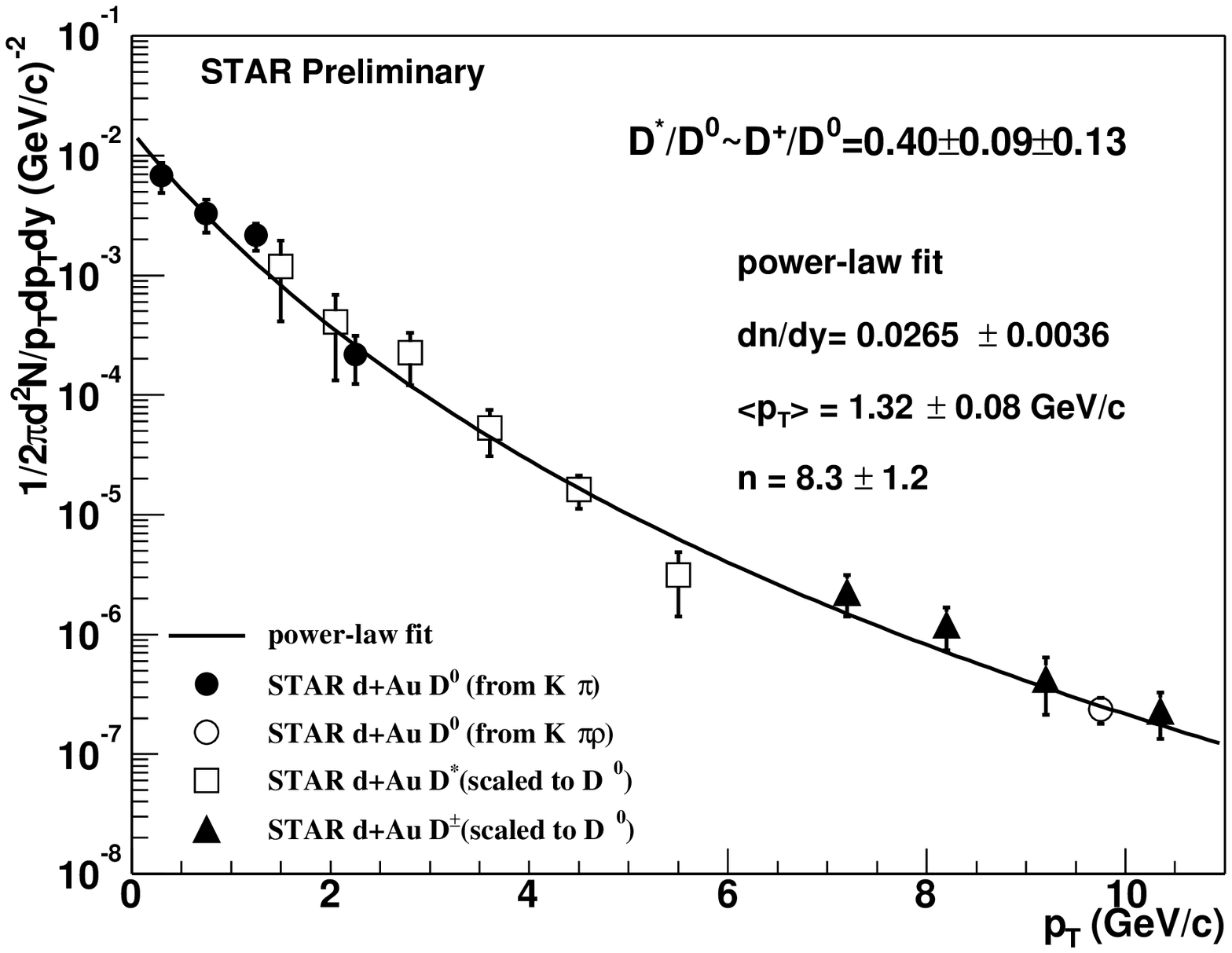,width=7.8cm, height=6. cm} 
\caption{Measured invariant multiplicity distributions for $D^0, D^{*+}$ and $D^{+}$ and a power-law fit to the data points. }
\label{dspectra}
\end{minipage}
\hspace{\fill}
\begin{minipage}[h]{76mm}
\epsfig{file=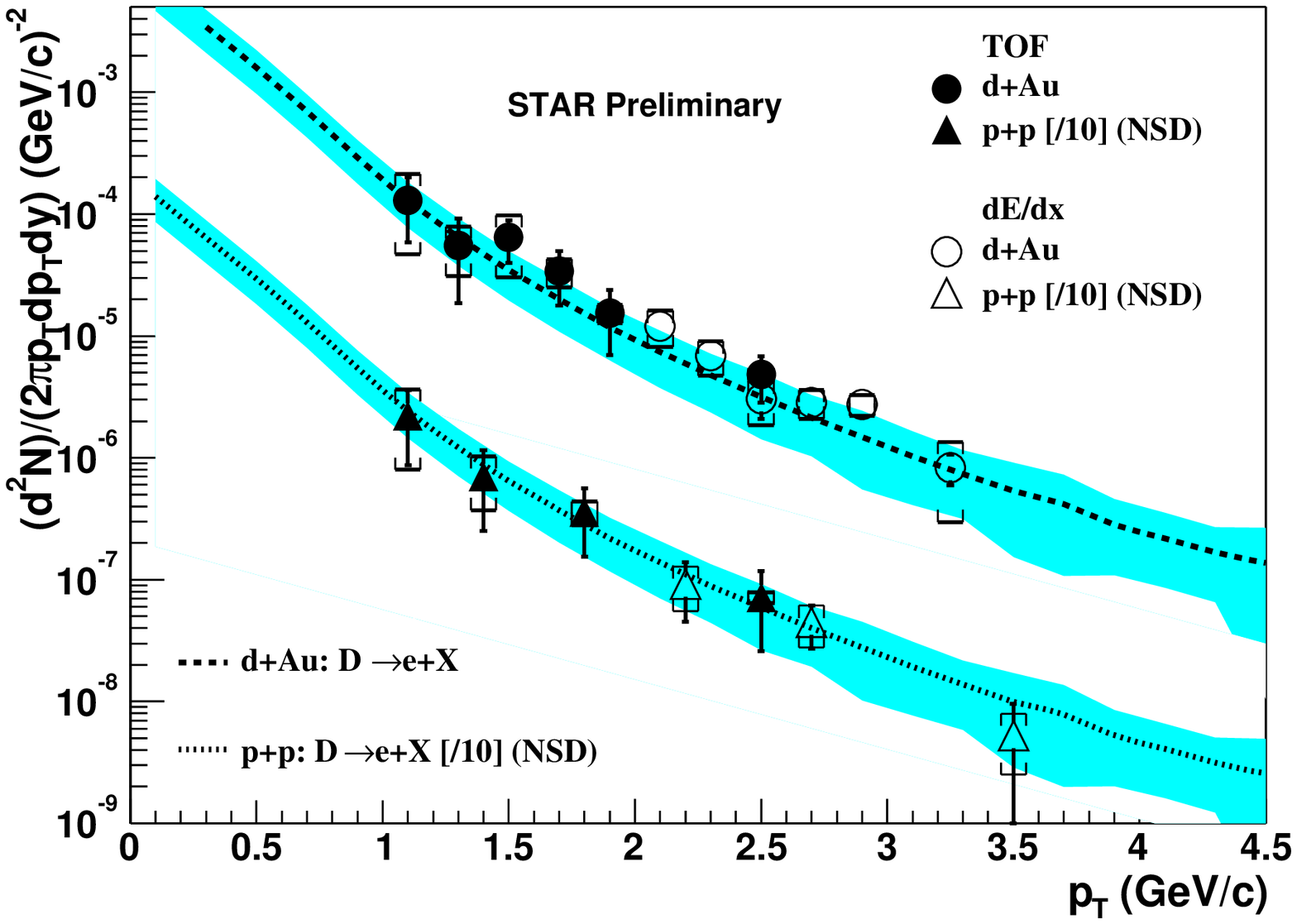,width=7.8cm,height=6.cm}
\caption{Consistency between directly measured open charm and the open charm measurements through single electrons. }
\label{electron}
\end{minipage}
\end{figure}

Fig.(\ref{dspectra}) shows the invariant multiplicity distributions of the measured open charms as a function of $p_{T}$ after the reconstruction efficiency correction, which is about 40\%-60\% (increasing
with $p_{T}$) except for $D^{*}$ at low $p_{T}$ (eg. 6\% at $p_{T}$=1.5 GeV/$c$ due to the low reconstruction efficiency of
the soft pions). The spectrum was  corrected for vertex and trigger efficiency. For $D^{0}\rightarrow K^+\pi^-$, a correction
was also done through a Monte Carlo simulation for the correlation around the $D^{0}$ mass 
due to the misidentification of the kaon and 
pion from the $D^{0}$ decay.  
Errors of the data points are statistical only and the data have been divided
by 2 assuming $\sigma(D)=\sigma(\bar{D})$. Overlap between $D^{0}$  and $D^{*}$
at low $p_{T}$ and between  the $D^{0}$  and $D^{+}$  at high $p_{T}$ allows us to 
compare the production cross
sections of different open charm species. It was found that the 
inclusive cross sections of  $D^{*}$ and $D^{+}$ are approximately equal. This observation
is  consistent with the measurements in~\cite{cdf,universal}. 

With the ratio of $D^*/D^0$(=$D^+/D^0$) as a free parameter we fit the all data points by a power-law function, 
$A(1+p_{T}/p_{0})^{-n}$,  where $A, p_{0}$,
and $n$ are parameters. From the fit, we obtained: $dn/dy(D^{0})=0.0265\pm 0.0036 (stat.)\pm 0.0071 (syst.) $ and $\langle p_{T}\rangle=1.32\pm0.08(stat.) \pm 0.16(syst.) $ GeV/$c$. 
It was also found that $D^{*}/D^{0}\approx D^{+}/D^{0}=0.40\pm 0.09(stat)\pm 0.13(sys)$. The $c\bar{c}$ cross section
can then be calculated by $\sigma_{c\bar{c}}=1.24(\sigma_{D^{0}}+\sigma_{D^{+}})$, where the factor 1.24 includes
the contribution from unmeasured $D_{s}$ and charm baryons ($\Lambda_{c}$ etc.)~\cite{universal}. Assuming number of binary
collision scaling, one could obtain  $\sigma_{D^{0}}$ for nucleon-nucleon collisions by 
$\sigma_{D^{0}}=4.6\times dn/dy(D^{0})\times \sigma_{inel}/N_{bin}$, where $\sigma_{inel}=42$~mb and $N_{bin}
=7.5$~\cite{starhighpt} 
are the inelastic cross section for a nucleon-nucleon collision and number of binary collisions in $d$+A collisions
at $\sqrt{s_{_{NN}}}$=200 GeV, respectively. The factor 4.6 accounts for the ratio of the $D^{0}$ multiplicity 
in the full rapidity region to $dn/dy(D^{0})$ at mid-rapidity and was obtained from a PYHTIA~\cite{pythia} 
calculation with  
parameters tuned to fit our measured spectrum~\cite{tunedpythia}. With the measured $dn/dy(D^{0})$ and $D^{+}/D^{0}$ ratio, 
the $\sigma_{c\bar{c}}$ for nucleon-nucleon collisions  is found to be $1.18 \pm 0.21(stat) \pm 0.39(sys)$~mb.

We also studied open charm production through measurements of single electrons from the open charm semileptonic decay.
Fig.(\ref{electron}) shows the electron spectra for $d$+Au and $p+p$ collisions at  $\sqrt{s_{_{NN}}}$=200 GeV, 
after background subtraction  in comparison with the expected electron spectrum
based on the decay of the STAR measured open charm spectrum (the bands). The background 
electrons are mainly from $\gamma$-conversion in the detector and from the $\pi^{0}$ and $\eta$ Dalitz decay.  
The single electrons were measured by both 
Time of Flight (TOF) (solid symbols) and TPC dE/dx (open symbols). The width of the bands reflects the uncertainties in the calculation.
For details about these measurements, see~\cite{lijuan}.  The 
open charm measurements through direct constructions of open charm signals and through the 
single electron measurement are consistent within STAR.

\begin{figure}[htb]
\begin{minipage}[h]{76mm}
\vspace{-.7cm}
\epsfig{file=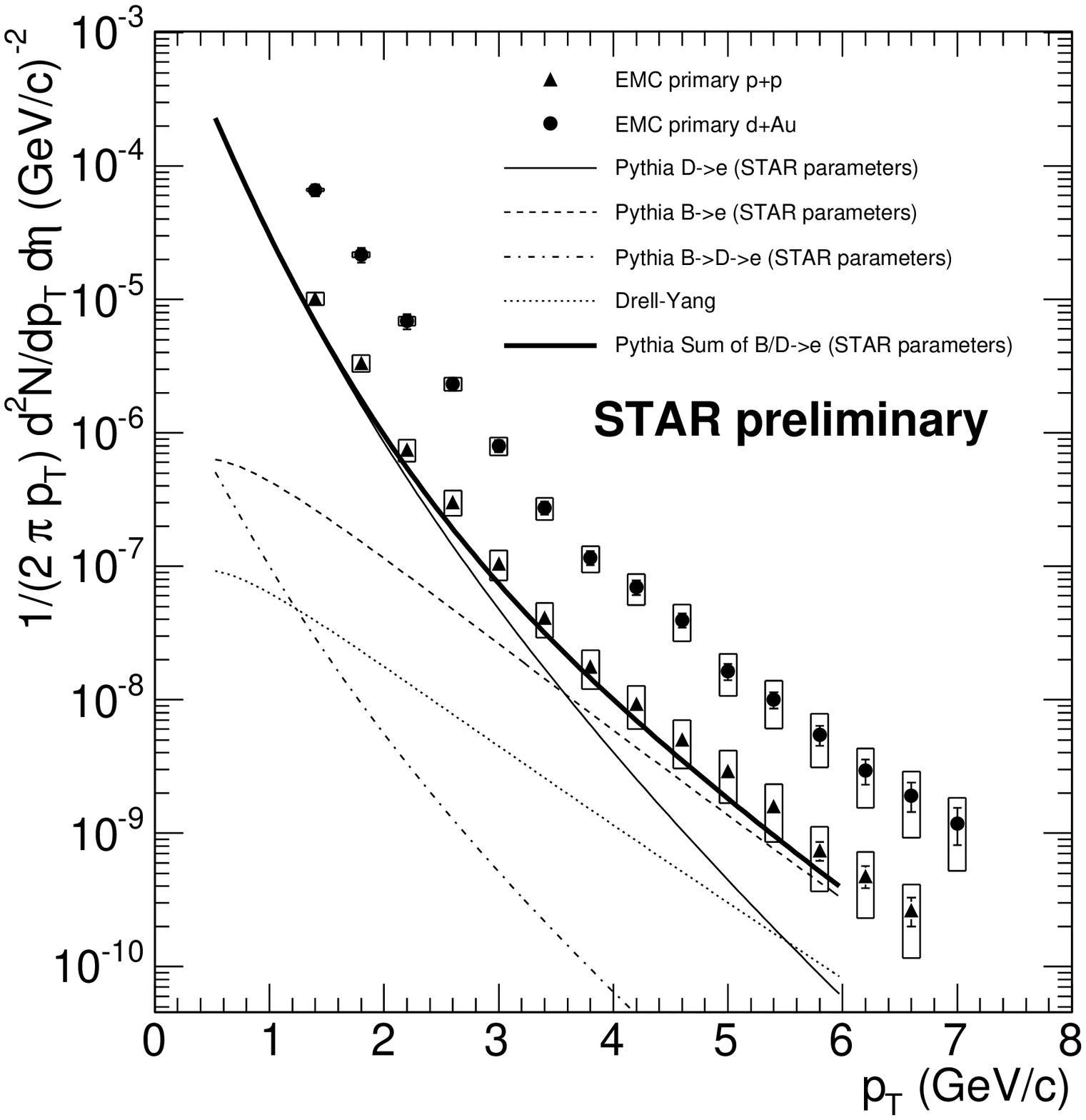,width=7.8cm, height=6. cm} 
\caption{Background subtracted single electron spectra in $p+p$ and $d$+Au collisions measured by STAR BEMC compared with a PYTHIA calculation.}
\label{emcfig}
\end{minipage}
\hspace{\fill}
\begin{minipage}[h]{76mm}
\vspace{-0.7 cm}
\epsfig{file=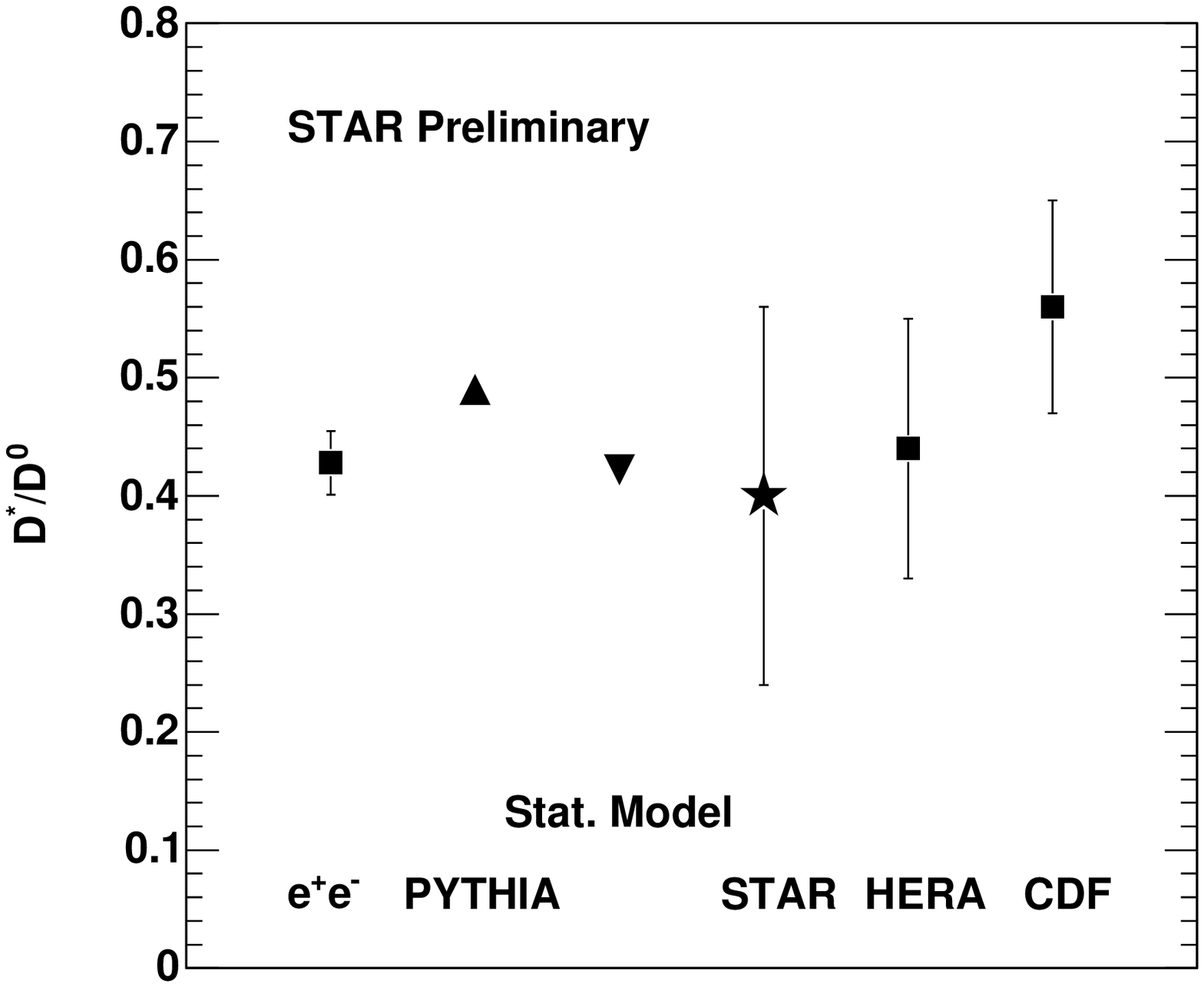,width=7.8cm,height=6. cm}
\caption{The STAR measured $D^{*}/D^{0}$ ratio in $d$+Au collisions compared with other experimental data and with theoretical predictions.}
\label{ratio}
\end{minipage}
\end{figure}
\begin{figure}[htb]
\begin{minipage}[h]{76mm}
\epsfig{file=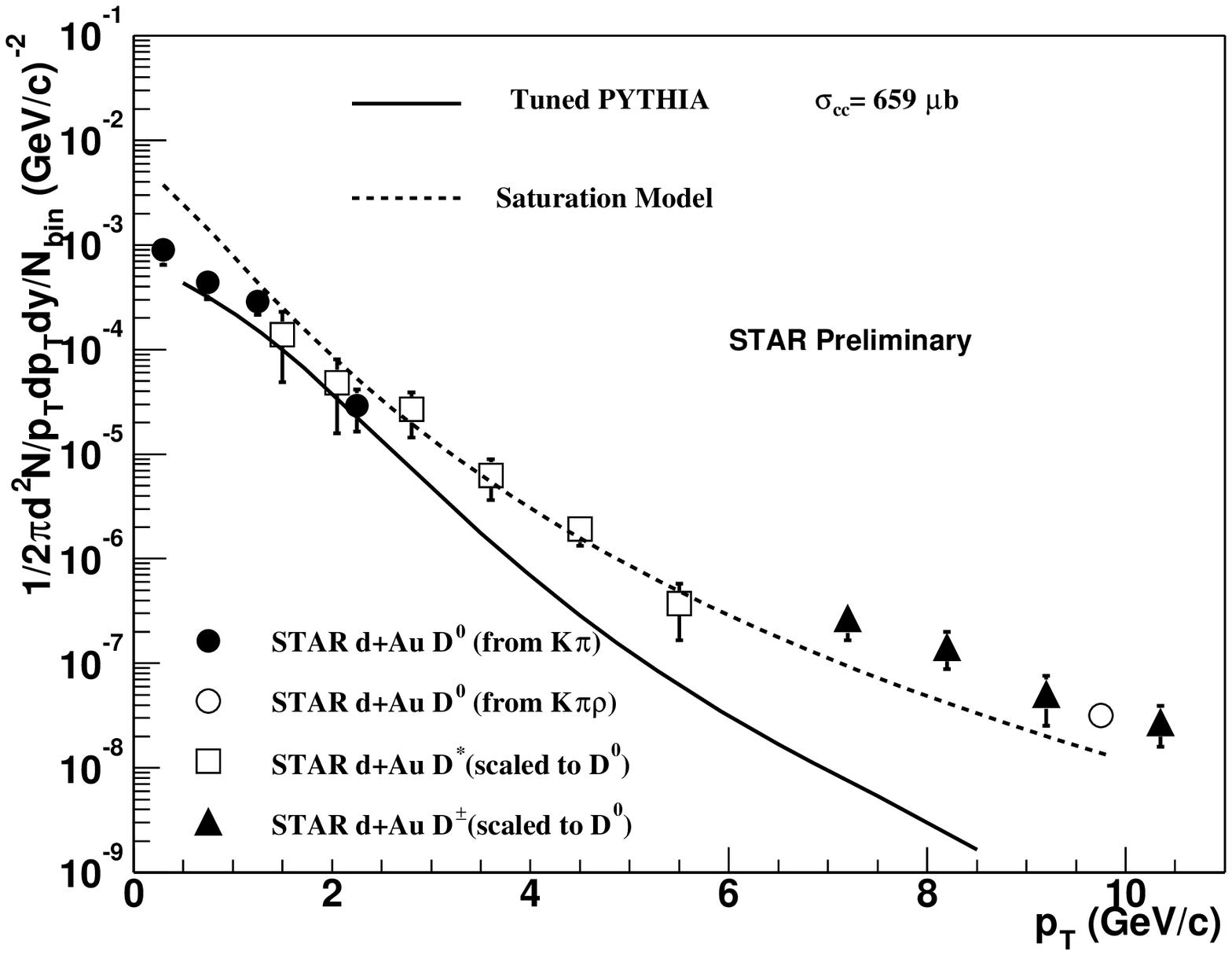,width=7.8cm, height=6. cm} 
\caption{The measured open charm spectrum compared with PYTHIA and the Saturation Model.}
\label{dima}
\end{minipage}
\hspace{\fill}
\begin{minipage}[h]{76mm}
\epsfig{file=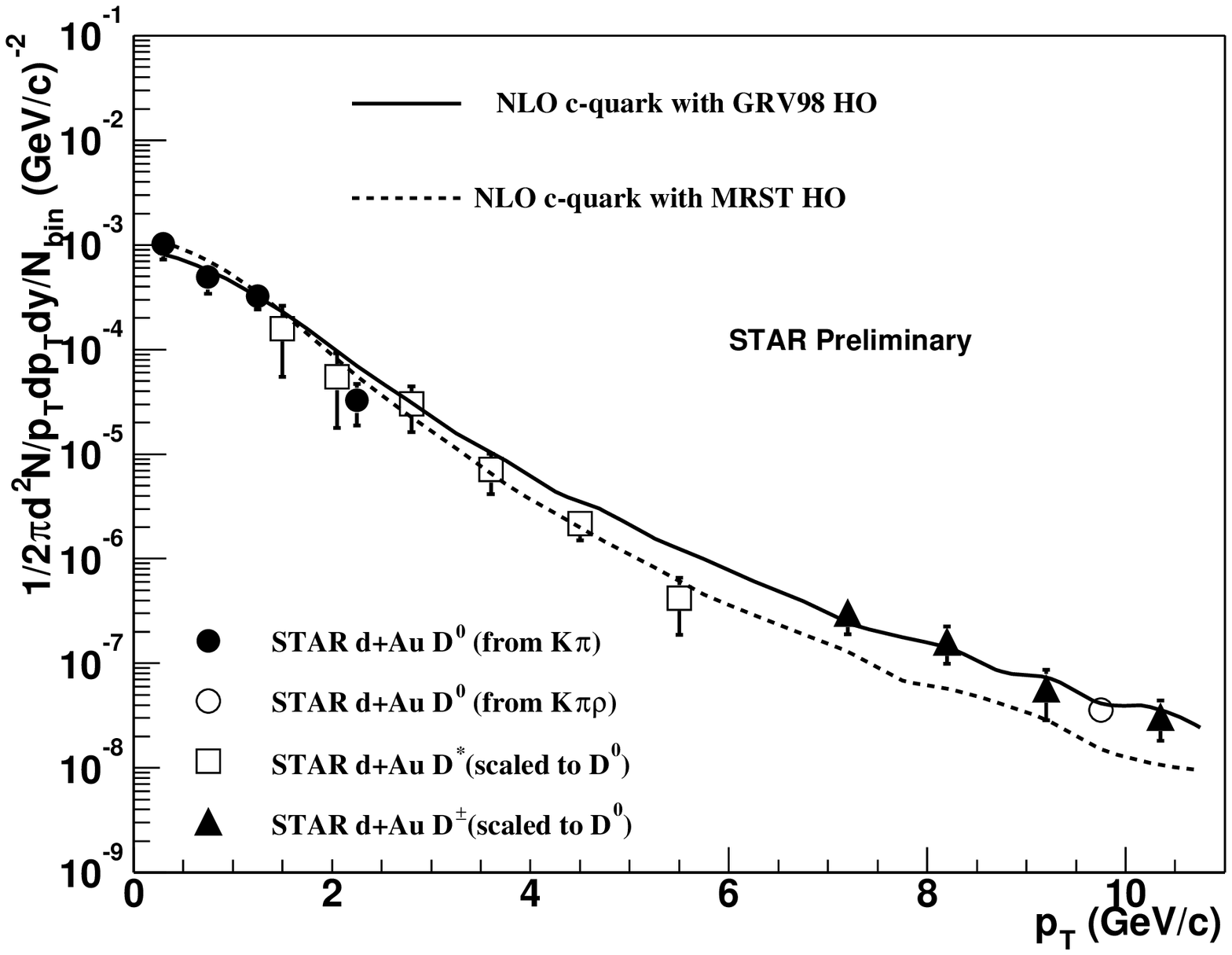,width=7.8cm,height=6. cm}
\caption{The measured open charm spectrum compared with NLO pQCD calculations for the charm quark. The theoretical
curves have been normalized to the measured cross section.}
\label{ramona}
\end{minipage}
\end{figure}
The single electrons were also measured  using the STAR Barrel Electromagnetic Calorimeter (EMC) up to 7.0 GeV/$c$. 
Fig.(\ref{emcfig}) gives the background subtracted electron spectra for $d$+Au (circles) and $p+p$ (triangles) 
collisions together with the calculations of PYTHIA with the modified parameters~\cite{tunedpythia}. The PYTHIA curves
shows that in the region of $p_{T}\gsim 4.0$ GeV/$c$ the contribution of B meson decays to electrons starts to overtake
that of D meson decays, which is the clear indication at RHIC that the bottom quark contributes significantly to the
measured high $p_{T}$ electrons. For details of this analysis, 
see~\cite{emc}.

In Fig.(\ref{ratio}), the STAR measured $D^*/D^{0}$ is compared with the previous measurements~\cite{cdf,universal} 
and with theoretical predictions~\cite{stat}. Within the experimental uncertainties the ratio shows little dependence
on collision system and energy.

The measured open charm spectrum is compared with theoretical predictions from the Saturation Model~\cite{dima} and
PYTHIA with parameters described in ref.~\cite{phenixele} in Fig.(\ref{dima}). Note that the open charm spectrum has been
scaled by $N_{bin}=7.5$. One sees that the measured open charm spectrum is much harder than the PYTHIA prediction.
However, the Saturation Model based on the $k_{T}$ factorization scheme does predict a harder spectrum in comparison
with one obtained from the LO pQCD calculation in PYTHIA. 

In Fig.(\ref{ramona}), the NLO pQCD 
predictions of the c-quark spectrum~\cite{ramona} are compared with the STAR open charm spectrum. 
The pQCD parameters in these calculations are  $m_c$ (charm quark mass)=1.2 GeV/$c^2$,  $\mu_{F}$(factorization scale)=$\mu_{R}$ 
(renormalization scale)=2 $m_{T}$ (transverse mass of the charm quark) for the  MRST HO PDF curve, and 
$m_c$=1.3 GeV/$c^2$,  $\mu_{F}$=$\mu_{R}$=$m_{T}$ for the  GRV98 HO PDF curve.  The theoretical curves have been normalized to the measured cross section though the $c\bar{c}$ cross sections from NLO
calculations  significantly underestimates the measured one (eg. by a factor of 3 with the MRST HO PDF).
It is interesting
that the measured spectrum shape of the open charm coincides with those of the c-quark, especially the one obtained
with GRV98 HO parton structure functions. Such a coincidence between the NLO c-quark spectrum and the measured open 
charm meson spectrum in charm hadroproduction has been observed in the fixed target experiments at low energies~\cite{wa92}. 
In order to fit with
data, the intrinsic$-k_{T}$ model had to be introduced to counter-balance the effect of c-quark hadronization 
through a fragmentation function of Peterson form~\cite{heavybig,wa92}. At RHIC, however, adding intrinsic$-k_{T}$  with a moderate
$<k_{T}^2>$ would not help to overcome the momentum
degradation of the fragmentation process  because the charm quark spectrum  
 is very broad. Such an observation
may indicate that the hadronization of a c-quark through the Peterson fragmentation function, which was established
in charm photoproduction~\cite{gammap} and in $e^+e^-$ collisions~\cite{ee}, is not suitable in case of charm hadroproduction.
Experimental data may suggest a fragmentation function more peaked towards the $x=1$ region.  A 
hadronization scheme of open charm production through quark recombination~\cite{rudy} was also proposed~\cite{recomb}. 

\subsection*{Acknowledgments}
We wish to acknowledge the many helpful discussions with R. Vogt.

\end{document}